\documentclass[prd,aps,showpacs,twocolumn,twoside]{revtex4-1}
\usepackage{amsfonts,amsmath,amssymb,bm,graphicx}
\usepackage[english]{babel}
\begin{document}

\title{Generation of hypermagnetic helicity and leptogenesis in early universe}

\author{V.B. Semikoz$^{a}$}
\email{semikoz@yandex.ru}
\author{A.Yu. Smirnov$^{a,b}$}
\email{smirnoff.alexandr@gmail.com}
\author{and}
\author{D.D. Sokoloff$^{a,c}$}
\email{sokoloff.dd@gmail.com}
\affiliation{$^{a}$Pushkov Institute of Terrestrial Magnetism, Ionosphere
and Radiowave Propagation (IZMIRAN), \\
142190 Troitsk, Moscow, Russia; \\
$^{b}$National University of Science and Technology "MISiS"
Moscow 119049, Russia, \\
$^{c}$Department of Physics, Moscow State University, Moscow, 119999 Russia}

%\date{\today}

\begin{abstract}
We study hypermagnetic helicity and lepton asymmetry evolution in plasma of the early Universe before the electroweak phase transition (EWPT) accounting for chirality flip processes via inverse Higgs decays and sphaleron transitions which violate the left lepton number and wash 
out the baryon asymmetry of the Universe (BAU). In the scenario where the right electron asymmetry supports the BAU alone through the conservation law $B/3 - L_{eR}=const$ at temperatures $T>T_{RL}\simeq 10~TeV$ the following universe cooling leads to the production of a non-zero left lepton (electrons and neutrinos) asymmetry. This is due to the Higgs decays becoming more faster when entering the equilibrium at $T=T_{RL}$ with the universe expansion, $\Gamma_{RL}\sim T> H\sim T^2$ , resulting in the parallel evolution of the right and the left electron asymmetries at $T<T_{RL}$ through the corresponding Abelian anomalies in SM in the presence of a seed hypermagnetic field. The hypermagnetic helicity evolution proceeds in a self-consistent way with the lepton asymmetry growth. The role of sphaleron transitions decreasing the left lepton number turns out to be negligible in given scenario. The hypermagnetic helicity plays a key role in lepto/baryogenesis in our scenario and the more hypermagnetic field is close to the maximum helical one the faster BAU grows up the observable value , $B_{obs}\sim 10^{-10}$.
\end{abstract}

% 97.60.Jd - Neutron stars
% 11.15.Yc - Chern-Simons gauge theory
% 25.30.Bf - Elastic electron scattering (nuclear physics)

%\keywords{Adler anomaly, magnetic field, chiral magnetic effect, parity violation}


%\begin{document}

\maketitle

\section{Introduction}

As a matter of fact, many celestial bodies contain magnetic field which is believed to be driven by 
electromagnetic induction, i.e. dynamo action. Magnetic fields are important for various physical processes 
including cosmic ray propagation, influence stellar (solar) activity, etc., while their origin is still an 
open problem in astrophysics and cosmology \cite{Betal12,Grasso:2000wj,BK}. It remains still unclear whether 
these magnetic field was at first created by battery effects in protogalaxies and then amplified by dynamo 
action up to the present-day strengths or the seed fields for dynamo action originate in magnetic fields which 
seem to be existed in the early Universe before recombination.  The first observational indications of the 
presence of cosmological magnetic fields (CMF) in the 
inter-galactic medium \cite{Neronov:2009gh,Neronov:1900zz} still do not exclude the first possibility however 
strongly support the latter option. 

The elaboration of the concept of magnetic field origin located in the very early Universe faced an obvious 
problem that we know very few about magnetic field at the very beginning of cosmological evolution. Each 
step of analysis shifts the problem for earlier and earlier stages of cosmological evolution and we are less 
and less informed about magnetic fields at these stages. At the first sight, it creates a rather unresolved 
problem for understanding of the topic, however, fortunately expansion of the Universe makes later stages 
relatively independent of the previous ones. Our aim here is to report such relative independence for one 
instructive quantity, namely magnetic helicity, which is achieved at the stage before the electroweak phase 
transition (EWPT) at the epoch when the magnetic helicity originates from the hepermagnetic one.

Remarkably, the problem occurs to be important for understanding of other fundamental physical property of 
contemporary Universe, i.e. substantial baryon-anti-baryon asymmetry known from everyday life and supported by 
astronomical observations. 

Specifically, we study lepto/baryogenesis in the early Universe plasma before EWPT in the presence of large-scale hypermagnetic fields (HMF) with an arbitrary hypermagnetic helicity density accounting for continuous HMF energy density spectra. This differs our approach from the simplified model with the maximum helicity density spectrum used in our previous works
\cite{Semikoz:2013xkc,Semikoz:2015wsa}. The choice of our scenario is motivated by the presence of the massless hypercharge fields $Y_{\mu}$ ($m_Y=0$) in hot electroweak plasma before EWPT that inevitably becomes a progenitor of the Maxwellian field $A_{\mu}$ after EWPT. One can also show that during EWPT of the first order supported by a strong HMF at $T\geq T_{EWPT}$ the hypermagnetic helicity fully converts to the magnetic one \cite{Akhmet'ev:2010ba}. 
The question how helical HMF can produce the observed baryon asymmetry of universe (BAU) as well as the following Maxwellian CMF evolution were studied in many papers \cite{Boyarsky:2011uy,Dvornikov:2011ey,Boyarsky:2012ex,Semikoz:2009ye,Semikoz:2012ka,Dvornikov:2012rk}. The Maxwellian chiral CMF evolution has been recently studied with the use of a non-uniform chiral anomaly in \cite{Boyarsky:2015faa}, applying  anomalous Maxwell equations for inhomogeneous chiral plasma in \cite{Shovkovy} and with the use of antisymmetric part of the photon polarization operator generated by a non-zero neutrino asymmetry in paper \cite{Dvornikov:2013bca}.The anomalous conversion of leptons to baryons during leptogenesis was studied also in paper \cite{Long:2013tha,Fujita:2016igl}, and the chiral change erasure via thermal fluctuations of magnetic helicity in the work \cite{Long:2016uez}.

Our work is organized as follows. In Sec. II we reproduce the scheme of our scenario developed in our previous papers \cite{Dvornikov:2011ey,Dvornikov:2012rk,Semikoz:2013xkc} starting from explanation of the current (\ref{current}) given by the Chern-Simons anomaly. In that Section we renew our approach considering an arbitrary initial HMF helicity. In subsection II.a we demonstrate in the case of a monochromatic spectrum how such a helicity tends to the maximal one due to the conservation law coming from the pair of self-consistent kinetic equations for spectra of the HMF helicity and HMF energy densities. Meaning an application to a realistic (say, Kolmogorov) spectrum we formulate in subsection II.b our initial conditions. Then in the next Sec. III we complete our kinetic equations by the leptogenesis for the first generation consisting of the singlet $e_R$ and the left doublet $L=(\nu_{eL} e_L)^T$ in external HMF's described by evolution equations derived in the previous Section. In Sec. IV we calculate  the HMF helicity density, and in Sec. V we calculate the baryon asymmetry. Finally in Sec. VI we discuss our results comparing them with some similar approaches in literature.
\vskip0.3cm

\section{Hypermagnetic helicity before EWPT}
The pseudovector current ${\bf J}$ induced in a seed hypermagnetic field ${\bf B}_Y=\nabla\times {\bf Y}$ enters the parity violation Chern-Simons (CS) term in the SM Lagrangian for the hypercharge field $Y_{\mu}$, $L_{CS}={\bf Y}\cdot{\bf J}$. This appears as the consequence of the polarization effect in a hot electroweak plasma \cite{Semikoz:2011tm,Dvornikov:2011ey},
\begin{equation}\label{current}
{\bf J}=\frac{g^{'2}}{4\pi^2}\left(\mu_{eR} + \frac{\mu_{eL}}{2}\right){\bf B}_Y,
\end{equation}
where $g^{'}=e/\cos\theta_W$ is the gauge coupling constant in SM, $\theta_W$ is Weinberg angle given by the experimental value $\sin^2\theta_W=0.23$; $\mu_{eR}$ and $\mu_{eL}$ are the chemical potentials for the right electron singlet $e_R$ and the left doublet $L=(\nu_{eL} e_L)^{\rm T}$, respectively.
Namely, this current which is additive to the ohmic one, ${\bf J}_{Ohm}=\sigma_{cond}({\bf E}_Y + {\bf V}\times {\bf B}_Y)$, leads to the instability of HMF in Faraday equation modified in SM.

We briefly mention the procedure in \cite{Semikoz:2011tm,Dvornikov:2011ey} leading to the sum of chemical potentials in Eq. (\ref{current}) when the CS term $L_{CS}={\bf Y}\cdot{\bf J}$ is derived in SM. The statistically averaged SM Lagrangian terms $f_R(g^{'})\langle\bar{e}_R\gamma_{\mu}e_R\rangle Y^{\mu} +f_L(g^{'})\langle\bar{e}_L\gamma_{\mu}e_L\rangle Y^{\mu} + f_L(g^{'})\langle\bar{\nu}_{eL}\gamma_{\mu}\nu_{eL}\rangle Y^{\mu}$, where $f_R(g^{'})=g^{'}y_R/2$, $f_L(g^{'})=g^{'}y_L/2$ play a role of "electric" charge associated to $U_Y(1)$, $y_R= -2$, $y_L=-1$ are hypercharges of the right-handed electron and the left-handed electron (neutrino) contribute, respectively, to the macroscopic 3-vector  ($\sim {\bf J}_{Ohm}$) and the 3-pseudovector ($\sim {\bf J}$) parts. The latter due to the phase volume used in the statistical average,\\ $f_{R,L}\langle\bar{l}_{R,L}\gamma_3\gamma_5l_{R,L}\rangle\sim f_{R,L}(g^{'})\sum_{n=0}^{\infty}[\mid f_{R,L}(g^{'})\mid B_Y/(2\pi)^2]\int_{-\infty}^{+\infty}dp_z (...)$, where $n=0,1,...$ is the Landau number, ${\bf B}_Y=(0,0,B_Y)$ is a seed HMF. Therefore, this current is proportional to the factor $-g^{'2}y_{R,L}^2$, or for the sum of lepton currents above one gets ${\bf J}\sim g^{'2}[4\mu_{eR} + \mu_{eL} + \mu_{\nu_{eL}}]{\bf B}_Y$. Since we should substitute $\mu_{eL}=\mu_{\nu_{eL}}$ for the left doublet $L=(\nu_{eL} e_L)^T$   the emergence of the sum $(\mu_{eR} + \mu_{eL}/2)$ in the current (\ref{current}) is obvious. Note that, as with the chiral magnetic effect \cite{Tashiro:2012mf} the current (\ref{current}) differs from zero for leptons (including neutrinos) at the main Landau level $n=0$ only, see details in \cite{Semikoz:2011tm}.

The Faraday equation derived in MHD from the Maxwell equation with the current (\ref{current}) added with the ohmic current ${\bf J}_{Ohm}=\sigma_{cond}({\bf E}_Y + {\bf v}\times {\bf B}_Y)$ reads \cite{footnote1}:
 \begin{equation}\label{Faraday}
\frac{\partial {\bf B}_Y}{\partial t}= \alpha_Y (t)\nabla\times {\bf B}_Y +  \eta_Y (t)\nabla^2{\bf B}_Y,
\end{equation}
where at temperatures $T_{RL}>T>T_{EW}$ the hypermagnetic helicity coefficient $\alpha_Y$ originates from the current (\ref{current}),
\begin{equation}\label{alpha}
\alpha_Y=\frac{g^{'2}}{4\pi^2\sigma_{cond}}\left(\mu_{eR} + \frac{\mu_{eL}}{2}\right),
\end{equation}
and $\eta_Y=(\sigma_{cond})^{-1}$ is the hypermagnetic diffusion coefficient, $\sigma_{cond}(T)\simeq 100T$ is the hot plasma conductivity.

From the Faraday equation one can obtain the evolution equations for the real binary products in the Fourier representation, $\partial_t\rho_{B_Y}(k,t)\sim [\dot{{\bf B}}_Y(k,t)\cdot{\bf B}_Y^*(k,t) + \dot{{\bf B}}_Y^*(k,t)\cdot{{\bf B}}_Y(k,t)] $ where $\rho_{B_Y}(t)=(2V)^{-1}\int(d^3k/(2\pi)^3)\mid {\bf B}_Y(k,t)\mid^2=\int dk \rho_{B_Y}(k,t))=B_Y^2(t)/2$ is the hypermagnetic energy density, and $\partial_t h_Y(k,t)\sim [\dot{{\bf Y}}(k,t)\cdot{\bf B}_Y^*(k,t) +{\bf Y}(k,t)\cdot\dot{{\bf B}}_Y^*(k,t)] $ for which $h_Y(t)=V^{-1}\int(d^3k/(2\pi)^3)[{\bf Y}(k,t)\cdot{\bf B}_Y^*(k,t)]=\int h_Y(k,t)dk$ is the hypermagnetic helicity density. We  use below conformal variables with the time $\eta=M_0/T$ where $M_0=M_{Pl}/ 1.66\sqrt{g^*}$, $M_{Pl}=1.2\times 10^{19}~{\rm GeV}$ is the Plank mass, $g^*=106.75$ is the number of relativistic degrees of freedom in the hot plasma before EWPT.

The general system of evolution equations for the spectra of the helicity density $\tilde{h}_Y(\tilde{k},\eta)$ and the energy density $\tilde{\rho}_{B_Y}(\tilde{k},\eta)$ obeying the inequality $\tilde{\rho}_{B_Y}(\tilde{k},\eta)\geq \tilde{k}\tilde{h}_Y(\tilde{k},\eta)/2$ \cite{Biskamp} reads \cite{Semikoz:2015wsa}:
\begin{eqnarray}\label{general}
\frac{d\tilde{h}_Y(\tilde{k},\eta)}{d\eta}=-\frac{2\tilde{k}^2}{\sigma_c}\tilde{h}_Y(\tilde{k},\eta) + &&\left(\frac{4\alpha^{'}[\xi_{eR}(\eta) + \xi_{eL}(\eta)/2]}{\pi\sigma_c}\right)\nonumber\\&&\times\tilde{\rho}_{B_Y}(\tilde{k},\eta),\nonumber\\
\frac{d\tilde{\rho}_{B_Y}(\tilde{k},\eta)}{d\eta}=-\frac{2\tilde{k}^2}{\sigma_c}\tilde{\rho}_{B_Y}(\tilde{k},\eta)+ &&\left(\frac{\alpha^{'}[\xi_{eR}(\eta) + \xi_{eL}(\eta)/2]}{\pi\sigma_c}\right)\nonumber\\&&\times\tilde{k}^2 \tilde{h}_Y(\tilde{k},\eta),
\end{eqnarray} 
where $\tilde{h}_Y(\tilde{k},\eta)=a^2h_Y(k,t)$ is the dimensionless HMF helicity density spectrum, $\tilde{k}=ak=const$ is the conformal wave number, $a=T^{-1}$ is the scale factor for FRW metric.
For the particular case of the maximum helicity $\tilde{h}_Y(\tilde{k},\eta)=2\tilde{\rho}_{B_Y}(\tilde{k},\eta)/\tilde{k}$ used in paper \cite{Semikoz:2013xkc} the system (\ref{general}) reads as the single equation:
\begin{eqnarray}\label{conformal}
\frac{d\tilde{h}_Y(\tilde{k},\eta)}{d\eta}=-\frac{2\tilde{k}^2\tilde{h}_Y(\tilde{k},\eta)}{\sigma_c} +&&\frac{2\alpha^{'}[\xi_{eR}(\eta) + \xi_{eL}(\eta)/2]\tilde{k}}{\pi\sigma_c}\nonumber\\&&\times\tilde{h}_Y(\tilde{k},\eta).
\end{eqnarray}
Here $\alpha^{'}=g^{'2}/4\pi$ is given by the SM coupling, $\sigma_c=\sigma_{cond}a=\sigma_{cond}/T\approx 100$ is the dimensionless plasma conductivity; $\xi_{eR}(\eta)=\mu_{eR}(T)/T$ and $\xi_{eL}(\eta)=\mu_{eL}(T)/T$ are the right and left electron asymmetry correspondingly.
\subsection{Tendency to the maximum helicity for small-scale HMF's}
Multiplying the first equation in the system (\ref{general}) by $(\tilde{k}^2/4)\tilde{h}_Y$, the second one by $\rho_{B_Y}$, and subtracting the first equation $\sim d(\tilde{h}_Y^2k^2/4)d\eta$ from the second one $\sim d\tilde{\rho}_{B_Y}^2/d\eta$ we cancel lepton asymmetry terms entering (\ref{general}) and obtain the simple differential equation:

\begin{equation}\label{simple}
\frac{{\rm d}}{{\rm d}\eta}\left(\tilde{\rho}_{B_Y}^2 - \frac{\tilde{h}_Y^2\tilde{k}^2}{4}\right)=-\frac{4\tilde{k}^2}{\sigma_c}\left(\tilde{\rho}_{B_Y}^2 - \frac{\tilde{h}_Y^2\tilde{k}^2}{4}\right).
\end{equation}
Accounting for large conformal times $\eta\to \eta_{EW}=7\times 10^{15}$ and the conductivity value $\sigma_c=100$, one finds that close to the EWPT the solution of Eq. (\ref{simple}),

\begin{eqnarray}\label{solution}
&&\tilde{\rho}_{B_Y}^2(\tilde{k},\eta) - \frac{\tilde{h}_Y^2(\tilde{k},\eta) \tilde{k}^2}{4}\nonumber\\&&=\left(\tilde{\rho}_{B_Y}^2(\tilde{k},\eta_0) - \frac{\tilde{h}_Y^2(\tilde{k},\eta_0) \tilde{k}^2}{4}\right)\exp \left(-\frac{4\tilde{k}^2}{\sigma_c}(\eta - \eta_0)\right)\nonumber\\
&&=\tilde{\rho}_Y^2(\tilde{k},\eta_0)(1 - q^2)\exp \left(-\frac{4\tilde{k}^2}{\sigma_c}(\eta - \eta_0)\right),
\end{eqnarray}
tends to the case of the maximum HMF helicity , $\tilde{h}_Y(\tilde{k},\eta)=2\tilde{\rho}_{B_Y}(\tilde{k},\eta)/\tilde{k}$, independently of the initial conditions at $T_0=T_{RL}$ given by the definition in Eq. (\ref{q-initial}) below.For example for large wave numbers of the order $\tilde{k}\sim 10^{-6}$, or for a small-scale HMF $\Lambda_{B_Y}=\tilde{k}^{-1}\simeq 10^6~T^{-1}$ at times $\eta\sim \eta_{EW}=7\cdot 10^{15}$ one gets in Eq. (\ref{solution}) the negligible exponential factor $\exp (-4\tilde{k}^2\eta_{EW}/\sigma_c)= \exp(-280)\approx 0$.
\begin{figure}
  \begin{center}
 % \subfigure[]
  {\label{PSaa}
  \includegraphics[scale=.35]{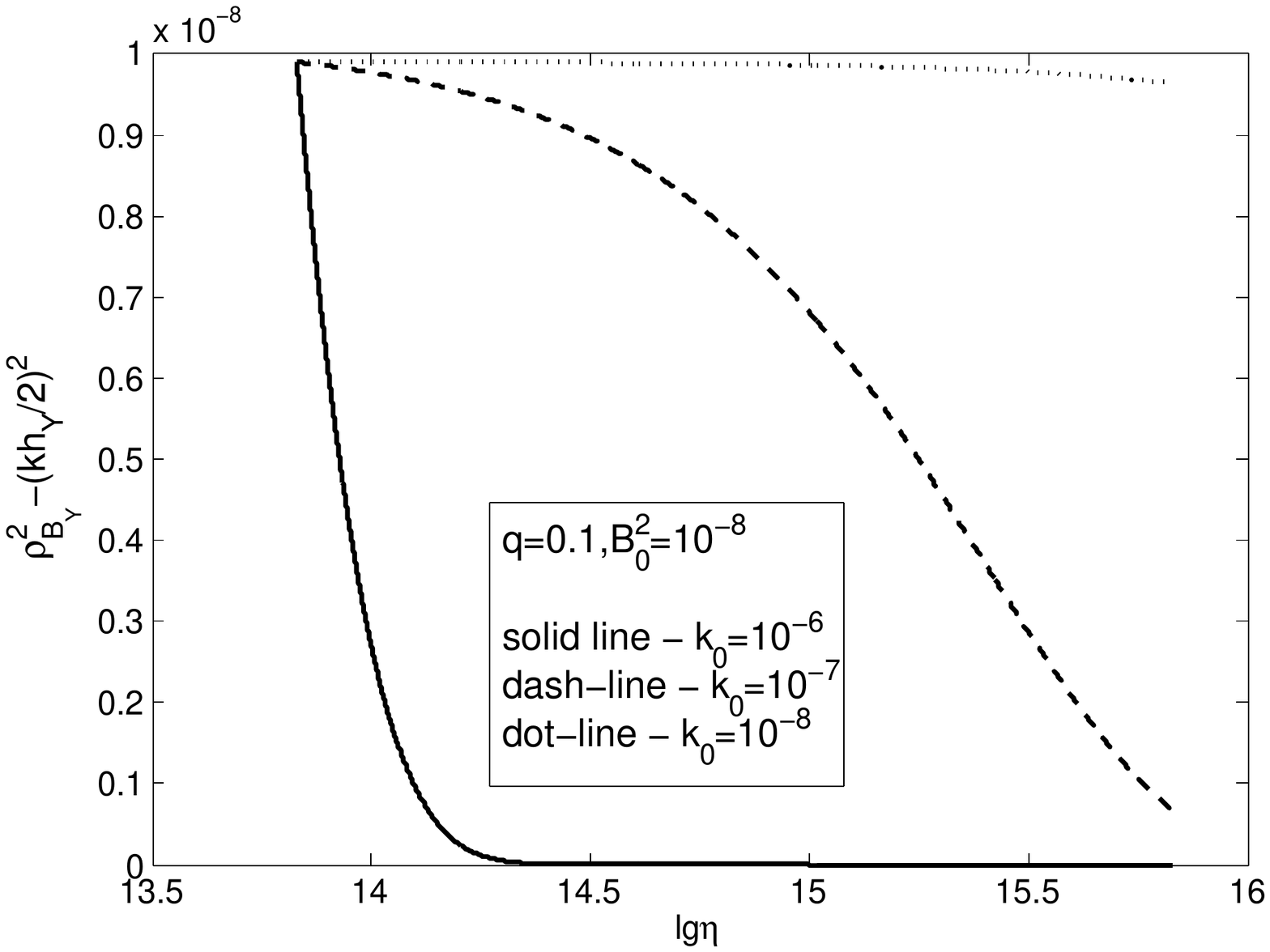}}
  %\hskip-.9cm
 % \subfigure[]
  {\label{PSbb}
  \includegraphics[scale=.35]{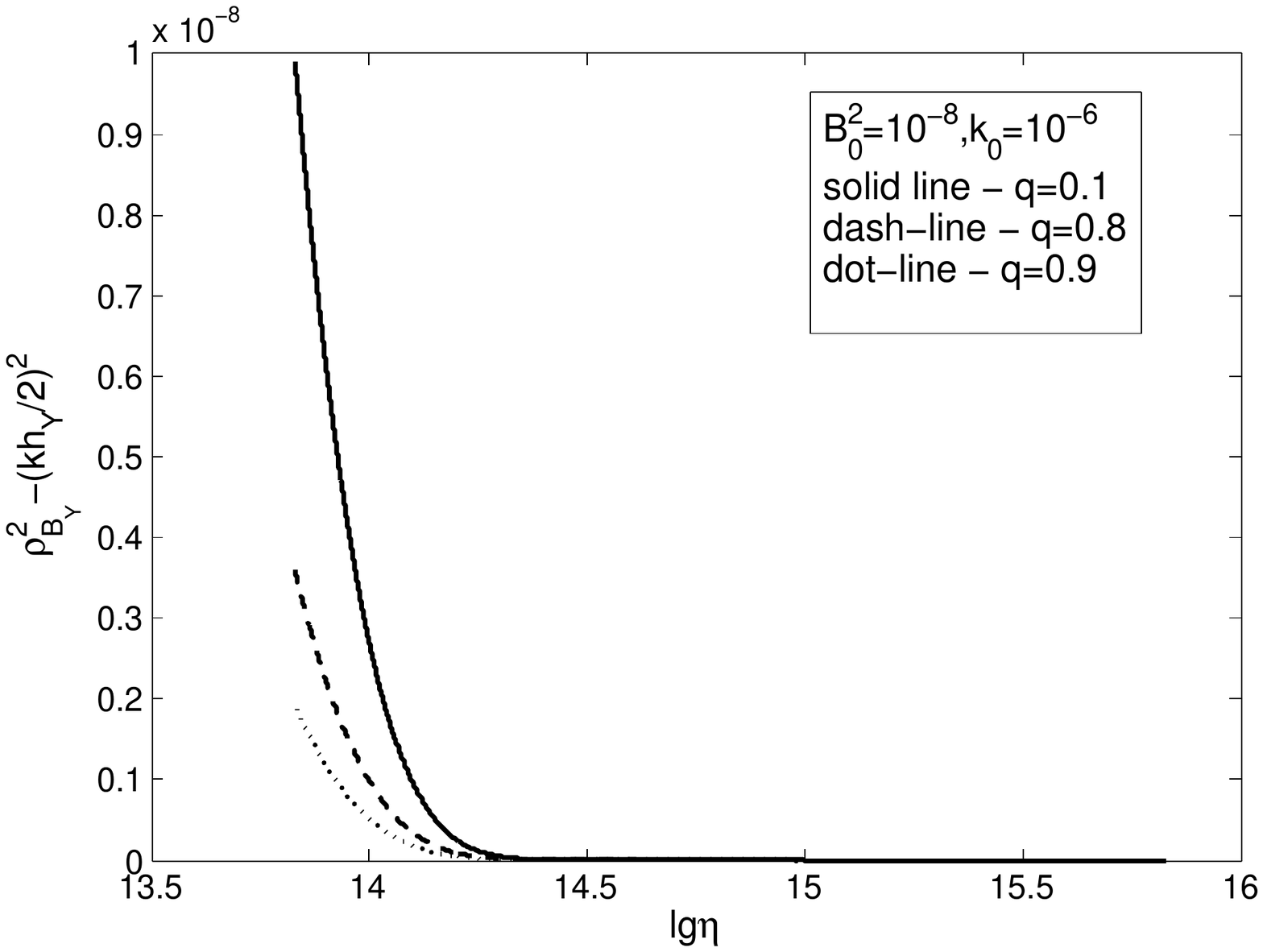}}
  \caption{The helicity growth to the maximum value given by  Eq. (\ref{solution}).
    Top panel corresponds to different monochromatic conformal $\tilde{k}=k/T=k_{0}$ and  the fixed factor $q=0.1$ in Eq (\ref{q-initial}).
    Bottom panel corresponds to different $q$ and same $\tilde{k}=k_{0}=10^{-6}$.The seed energy density (here doubled) $B_0^2=10^{-8}=2\rho_{B_Y}^{(0)}$, corresponds to the strength of conformal (dimensionless) HMF $\tilde{B}_Y^{(0)}=B_Y^{(0)}/T^2$ squared.
  \label{fig:BAU1}}
  \end{center}
\end{figure}

One can see in the top panel  that for a large scale $\Lambda_{B_Y}=k_0^{-1}$ corresponding to $k_0=10^{-8}$ (dotted line) the helicity density does not reach the maximum helicity condition $h_Y\to 2\rho_{B_Y}/k_0$ to the end of its evolution at $\eta_{EW}=7\times 10^{15}$. This means that in the general case of a continuous spectrum, e.g. for the Kolmogorov one $\rho_{B_Y}(\tilde{k},\eta_0)\sim k^{-5/3}$ \cite{footnote4}
%\footnote{We stress that we are dealing with helical fields and its spectrum can be rather more complicated than the classical %Kolmogorov spectrum \cite{Kahniashvili:2012uj}.}, 
,it would be difficult to expect an implementation of that condition during a short time, $\eta_0\leq \eta\leq \eta_{EW}$.

There is also a danger to spread the region of continuous spectra, $0\leq \tilde{k}\leq \tilde{k}_{max}$, to a larger value $\tilde{k}_{max}$. Firstly, we neglected the small-scale fluid velocity correlation lengths, $\lambda_v\ll \Lambda_{B_Y}$, to avoid the necessity to add the Navier-Stokes equation for the matter fluid ($\sim {\bf v}$) to our Faraday equation (\ref{Faraday}). Secondly, as it was shown in previous papers \cite{Dvornikov:2011ey,Dvornikov:2012rk} for a monochromatic helicity spectrum, for large wave numbers $k_0$ the baryon asymmetry grows too much before EWPT exceeding significantly the observable BAU value, $B_{obs}=10^{-10}$.

Substituting HMF energy density spectrum $\tilde{\rho}_{B_Y}(\tilde{k},\eta)$ from the relation (\ref{solution}) into the first equation for $d\tilde{h}_Y/d\eta$ in the system (\ref{general}) one can easily get its solution as
\begin{eqnarray}\label{helicity_integral}
&&\tilde{h}_Y(\tilde{k},\eta)=\frac{2\tilde{\rho}_{B_Y}(\tilde{k},\eta_0)}{\tilde{k}}\Bigl[\sinh\left(\frac{2\alpha^{'}\tilde{k}}{\pi\sigma_c}\int_{\eta_0}^{\eta}\Xi (\eta^{'})d\eta^{'}\right) \nonumber\\&& +q\cosh\left(\frac{2\alpha^{'}\tilde{k}}{\pi\sigma_c}\int_{\eta_0}^{\eta}\Xi (\eta^{'})d\eta^{'}\right)\Bigr]e^{\frac{-2\tilde{k}^2(\eta - \eta_0)}{\sigma_c}}
\end{eqnarray}

where the integrand in argument of hyperbolic functions, $\Xi(\eta)=\xi_{eR}(\eta) + \xi_{eL}(\eta)/2$, should be found from the kinetic equations for asymmetries $\xi_{eR}(\eta)$, $\xi_{eL}(\eta)$, see in Sec.~3.

\subsection{Initial conditions}
We choose the general initial condition for the hypermagnetic helicity density obeying the known inequality for spectra $h(t,k)\leq 2\rho_B(t,k)/k$ \cite{Biskamp} and consistent with Eq. (\ref{helicity_integral}):
\begin{equation}\label{q-initial}
\tilde{h}_Y(\tilde{k},\eta_0)=q\left(\frac{2\tilde{\rho}_{B_Y}(\tilde{k},\eta_0)}{\tilde{k}}\right),~~~~~~0\leq q \leq 1,
\end{equation}
where the cases $q=0$ ($q=1$) corresponds to the non-helical initial HMF (fully helical HMF). Note that the case
$q=1$ with the system of master equations (\ref{general}) reduced to the single evolution equation (\ref{conformal}) was already studied in Ref. \cite{Semikoz:2013xkc}. The continuous initial HMF energy density spectrum $\tilde{\rho}_{B_Y}(\tilde{k},\eta_0)$ is given by the index
$n_{B_Y}$,
\begin{equation}\label{energy}
\tilde{\rho}_{B_Y}(\tilde{k},\eta_0)= A\tilde{k}^{ n_{B_Y}},
\end{equation}
where the normalization constant $A$,
\begin{equation}\label{constant}
A=\frac{(1 + n_{B_Y})(\tilde{B}_0^Y)^2}{2(\tilde{k}_{max}^{1 + n_{B_Y}} - \tilde{k}_{min}^{1 + n_B})},
\end{equation}
 is given by a seed HMF $\tilde{B}_0^Y=\sqrt{2\tilde{\rho}_{B_Y}^{(0)}}$ with the fixed  $\tilde{\rho}_{B_Y}^{(0)}=10^{-8}$; $\tilde{k}_{min}(\eta_0)=(T_0\cdot l_H)^{-1}=T_0/M_0=\eta_0^{-1}$ corresponds to the largest scale $\Lambda_{B_Y}=k_{\min}^{-1}=l_H(\eta_0)$ \cite{footnote2} and $\tilde{k}_{max}$ is an arbitrary wave number parameter constrained as discussed previously.

In Eq. (\ref{energy}) we choose $n_{B_Y}= - 5/3$ for the Kolmogorov spectrum, while other models of continuous spectra are possible, e.g. the Kazantsev spectrum with $n_{B_Y}= -1/2$, or the white noise case $n_{B_Y}=0$.

Below in the self-consistent evolution equations for  lepton asymmetries (\ref{right}), (\ref{left}) we choose the initial values $\xi_{eR}(\eta_0)= 10^{-14}$ and $\xi_{eL}(\eta_0)=0$ in our scenario. Here the left lepton asymmetry is absent at high temperatures $T> T_0=T_{RL}$ for the conformal times $\eta<\eta_0$ where $\eta_0=M_0/T_{RL}= 7\times 10^{13}$ is the initial time for $T_{RL}=10~{\rm TeV}$.
Let us remind how the leptogenesis looks in our model \cite{Semikoz:2013xkc}.

\section{Leptogenesis in hypermagnetic fields}
 For simplicity we consider inverse Higgs decays only or we neglect the Higgs boson asymmetry, $\mu_0=0$.
The system of kinetic equations for leptons accounting for Abelian anomalies for right electrons and left electrons (neutrinos), $\partial_{\mu}j^{\mu}_{R,L}=\pm g^{'2}y_{R,L}^2{\bf E}_Y\cdot{\bf B}_Y/16\pi^2$, inverse Higgs decays and sphaleron transitions as well, takes the form \cite{Semikoz:2013xkc}:
\begin{eqnarray}\label{system} 
  \frac{{\rm d}L_{e_\mathrm{R}}}{\rm dt} = 
  \frac{g'^2}{4\pi^2s} (\mathbf{E}_\mathrm{Y} \cdot \mathbf{B}_\mathrm{Y}) + &&
  2\Gamma_\mathrm{RL}
  \left\{
    L_{e_\mathrm{L}}-L_{e_\mathrm{R}}\right\},
\nonumber \\
  \frac{{\rm d}L_{e_\mathrm{L}}}{\rm dt} = 
  -\frac{g'^2}{16\pi^2s}(\mathbf{E}_\mathrm{Y} \cdot \mathbf{B}_\mathrm{Y}) +&&
  \Gamma_\mathrm{RL}
  \left\{
    L_{e_\mathrm{R}} - L_{e_\mathrm{L}}\right\}\nonumber\\&&-\left(\frac{\Gamma_{sph}T}{2}\right)L_{e_\mathrm{L}}.
 \end{eqnarray}
Here $L_b=(n_b - n_{\bar{b}})/s\approx T^3\xi_b/6s$ is the lepton number, $b=e_\mathrm{R},e_\mathrm{L}, \nu_e^\mathrm{L}$, $s=2\pi^2g^*T^3/45$ is the entropy density, and $g^*=106.75$ is the number of relativistic degrees of freedom. The factor=2 in the first line takes into account the equivalent reaction branches, $e_R\bar{e}_L\to \tilde{\varphi}^{(0)}$ and $e_R\bar{\nu}_{e^\mathrm{L}}\to \varphi^{{(-)}}$; $\Gamma_{RL}$ is the chirality flip rate. Of course, for the left doublet $L_e^T=(\nu_{e^\mathrm{L}},e_L)$ kinetic equation for neutrino number is excess because $L_{e_\mathrm{L}}=L_{\nu_e^\mathrm{L}}$. Then $\Gamma_{sph}=C\alpha_W^5=C(3.2\times 10^{-8})$ is the dimensionless probability  of sphaleron transitions decreasing the left lepton numbers and therefore washing out baryon asymmetry of universe (BAU). It is given by the $SU(2)_W$ constant $\alpha_W=g^2/4\pi=1/137\sin^2\theta_W=3.17\times 10^{-2}$ where $g=e/\sin \theta_{W}$ is the gauge coupling  in SM and the constant $C\simeq 25$ is estimated through lattice calculations (see, e.g., the chapter 11 in the book \cite{Rubakov}).

In conformal variables after integration of the  system (\ref{system}) over volume $\int d^3x(...)/V$, transferring to the Fourier components for hypercharge fields the kinetic equations (\ref{system}) take the form
 \begin{eqnarray}\label{right}
\frac{d\xi_{eR}(\eta)}{d\eta}=- \frac{3\alpha^{'}}{\pi}\int &&d\tilde{k} \frac{d\tilde{h}_Y(\tilde{k},\eta)}{d\eta}  \nonumber\\&&-\Gamma\Bigl[\xi_{eR}(\eta) -\xi_{eL}(\eta)\Bigr],
\end{eqnarray}
\begin{eqnarray}\label{left}
\frac{d\xi_{eL}(\eta)}{d\eta}&&=+ \frac{3\alpha^{'}}{4\pi}\int d\tilde{k} \frac{d\tilde{h}_Y(\tilde{k},\eta)}{d\eta} \nonumber\\&&-\frac{\Gamma (\eta)}{2}\Bigl[\xi_{eL}(\eta) - \xi_{eR}(\eta)\Bigr] - \frac{\Gamma_{sph}}{2}\xi_{eL}(\eta),
\end{eqnarray}
where 
\begin{equation}\label{rate}
\Gamma(\eta)=\left(\frac{242}{\eta_{EW}}\right)\left[1 - \left(\frac{\eta}{\eta_{EW}}\right)^2\right],~~~\eta_{RL}<\eta <\eta_{EW}
\end{equation}
is the dimensionless chirality flip rate $\Gamma=2a\Gamma_{RL}$ \cite{Dvornikov:2011ey,Campbell:1992jd} ,
$\eta_{EW}=M_0/T_{EW}=7\times 10^{15}$ is the EWPT time at $T_{EW}=100~GeV$. 

\begin{figure}
  \begin{center}
%\centering
  %\subfigure[]
  {\label{PSa}
  \includegraphics[scale=0.36]{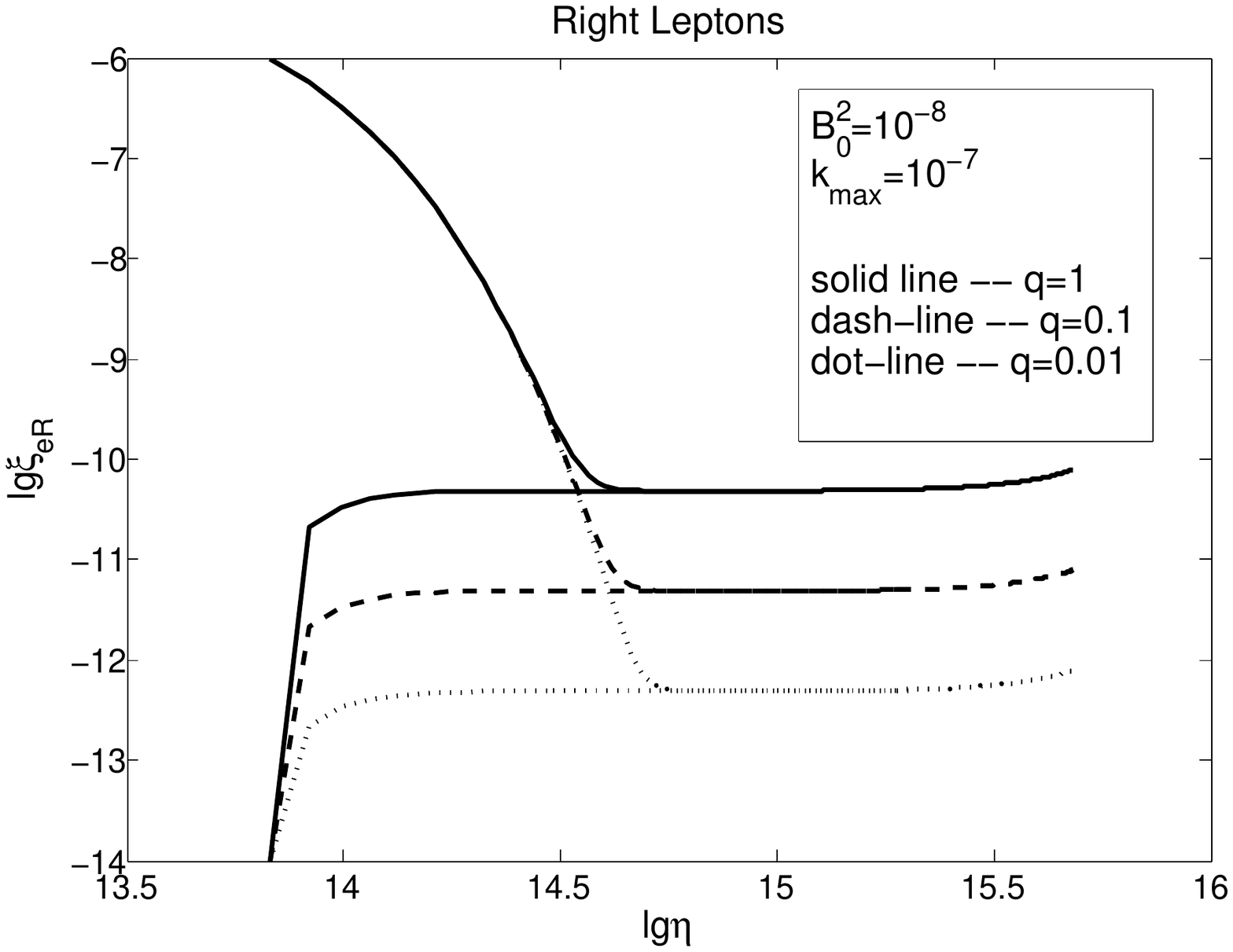}}
  %\subfigure[]
  {\label{PSb}
  \includegraphics[scale=0.36]{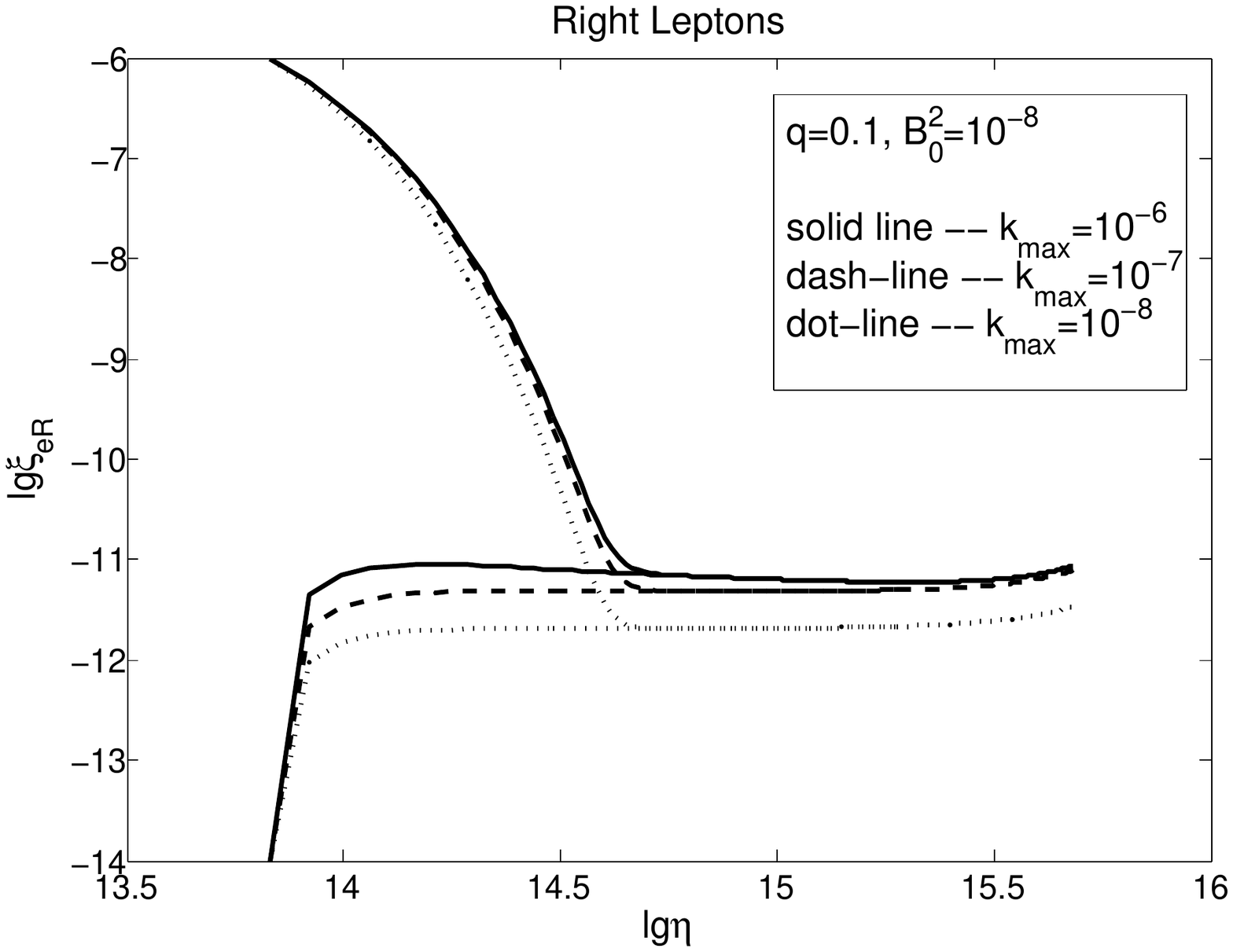}}
  
  \caption{The evolution of the right electron asymmetry $\xi_{eR}(\eta)$ for the continuous initial Kolmogorov spectrum (\ref{energy}). 
  Top panel corresponds to different factors $q$ in Eq.~(\ref{q-initial}), different initial value $\xi_{eR}(\eta_{0})$: $\xi_{eR}(\eta_0)=10^{-6}$ and $\xi_{eR}(\eta_0)=10^{-14}$ and same conformal $\tilde{k}_{max}=10^{-6}=k_{max}/T$ denoted as $k_{max}$ .
  Bottom panel corresponds to different $k_{max}$, different initial value $\xi_{eR}(\eta_{0})$ and same $q=0.1$. The seed energy density (here doubled) $B_0^2=10^{-8}=2\rho_{B_Y}^{(0)}$ is chosen as in Fig. 1.
  \label{fig:RLeptons}}
  \end{center}
\end{figure}

Here we solve self-consistent kinetic equations (\ref{right}), (\ref{left}) substituting the derivative $d\tilde{h}_Y(\tilde{k},\eta)/d\eta$ from the the formal solution of system (\ref{general}) for spectra $\tilde{\rho}_{B_Y}(\tilde{k},\eta)$ and 
$\tilde{h}_Y(\tilde{k},\eta)$ given by Eq. (\ref{helicity_integral}).

In Fig.\ref{fig:RLeptons} we show the evolution of the right lepton asymmetry $\xi_{eR}(\eta)$ found from the system of self-consistent eqs. (\ref{right}), (\ref{left}) that can help us to interpret the BAU evolution seen in Fig 4 .  Note that the left lepton asymmetry $\xi_{eL}$
has a much smaller value, $\xi_{eL}\ll \xi_{eR}$ , first, due to sphaleron transitions which reduce $L_{eL}$, second, in our scenario we choose the initial conditions $\xi_{eL}(\eta_0)=0$, $\xi_{eR}(\eta_0)\neq 0$ for which $\xi_{eL}$ does not have time to grow down to the EWPT time $\eta_{EW}$.
Indeed, assuming the saturation limit $\partial_t\xi_{eR}=\partial_t\xi_{eL}\approx 0$, multiplying (\ref{left}) by the factor four and adding that with (\ref{right}) one finds the inequality
\begin{equation}\label{inequality}
\xi_{eL}=\frac{\Gamma\xi_{eR}}{\Gamma + 2\Gamma_{sph}}\ll \xi_{eR},
\end{equation} 
where $\Gamma_{sph}\gg \Gamma$.  A discussion why $\xi_{eR}(\eta)$ growing due to the Abelian anomaly  tends to  $\xi_{eR}(\eta)\approx constant$ is done in Ref. \cite{Semikoz:2013xkc} where {\it in the case of the monochromatic helicity density spectrum} authors showed the independence of such saturation level from  a chosen initial condition $\xi_{eR}(\eta_0)=10^{-10}$ or $\xi_{eR}(\eta_0)=10^{-4}$. The similar independence of the saturation values $\xi_{eR}\approx const$ is seen here for {\it continuous (Kolmogorov) helicity density spectrum} when comparing curves  for same free parameters $q$ in (\ref{q-initial}) ( running $q=0.01,~0.1,~1$ ), which start from different initial asymmetries $\xi_{eR}(\eta_0)$ in the top panel in Fig.~\ref{fig:RLeptons}, and curves for the same $\tilde{k}_{max}$ (running $\tilde{k}_{max}=10^{-8}\div 10^{-6}$) and fixed $q=0.1$ in the bottom panel. Then the growing tail in the right lepton asymmetry evolution in Fig.\ref{fig:RLeptons} is given by the vanishing rate of Higgs decays (\ref{rate}) and leads to the additional growth of the BAU in Fig 4 when $\eta\to \eta_{EW}$.

\vskip0.3cm
\section{HMF helicity density evolution for Kolmogorov spectrum}
In this Section we calculate the HMF helicity density for its arbitrary initial level given by the parameter $q\leq 1$.
In paper \cite{Semikoz:2013xkc} using single Eq.(\ref{conformal}) we assumed a fully helical field from the beginning while such assumption may be not the case ( see e.g., in \cite{Tevzadze:2012kk}).

\begin{figure}
  \begin{center}
%\centering
  %\subfigure[]
  {\label{HDa}
  \includegraphics[scale=0.35]{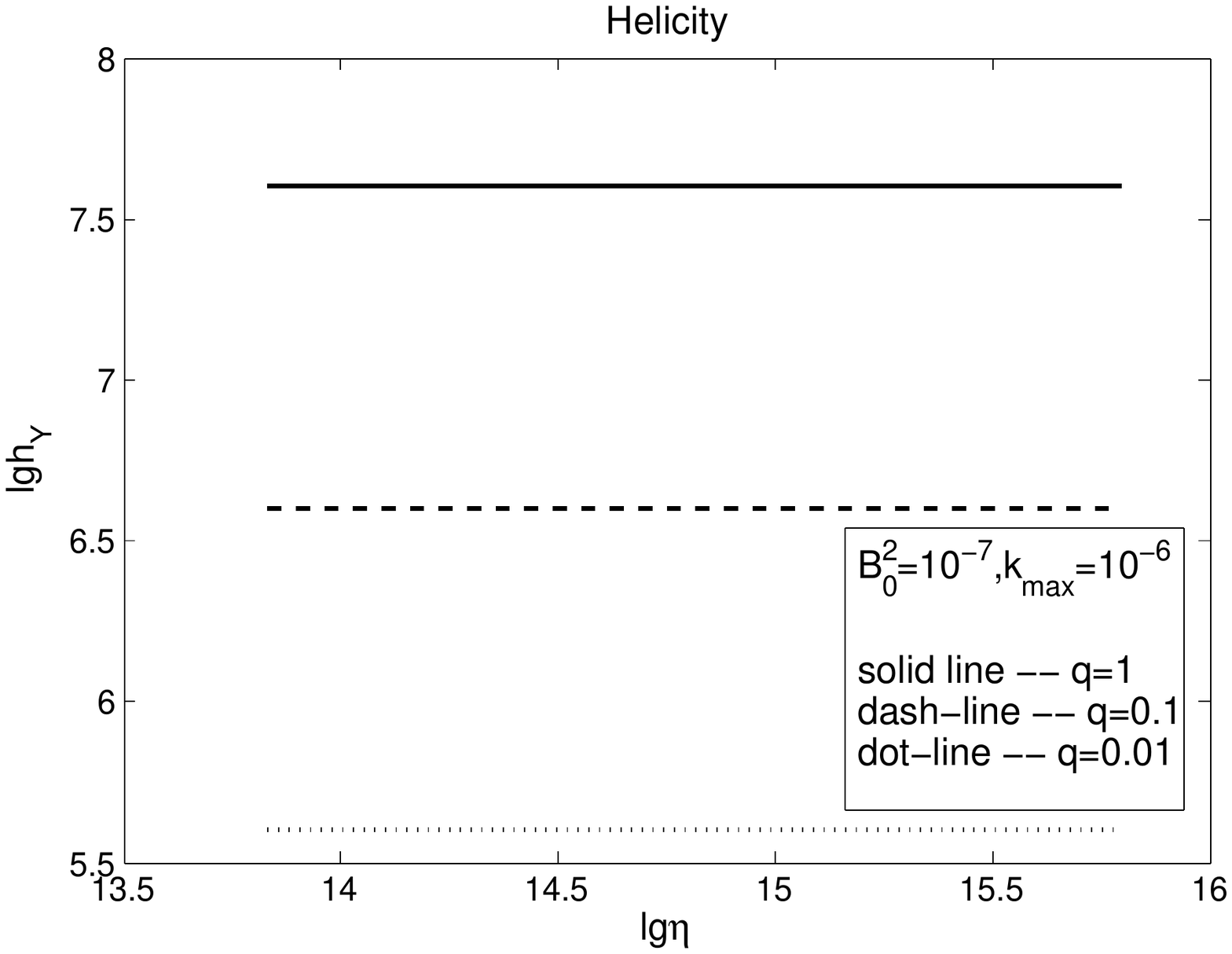}}
  %\subfigure[]
  {\label{HDb}
  \includegraphics[scale=0.35]{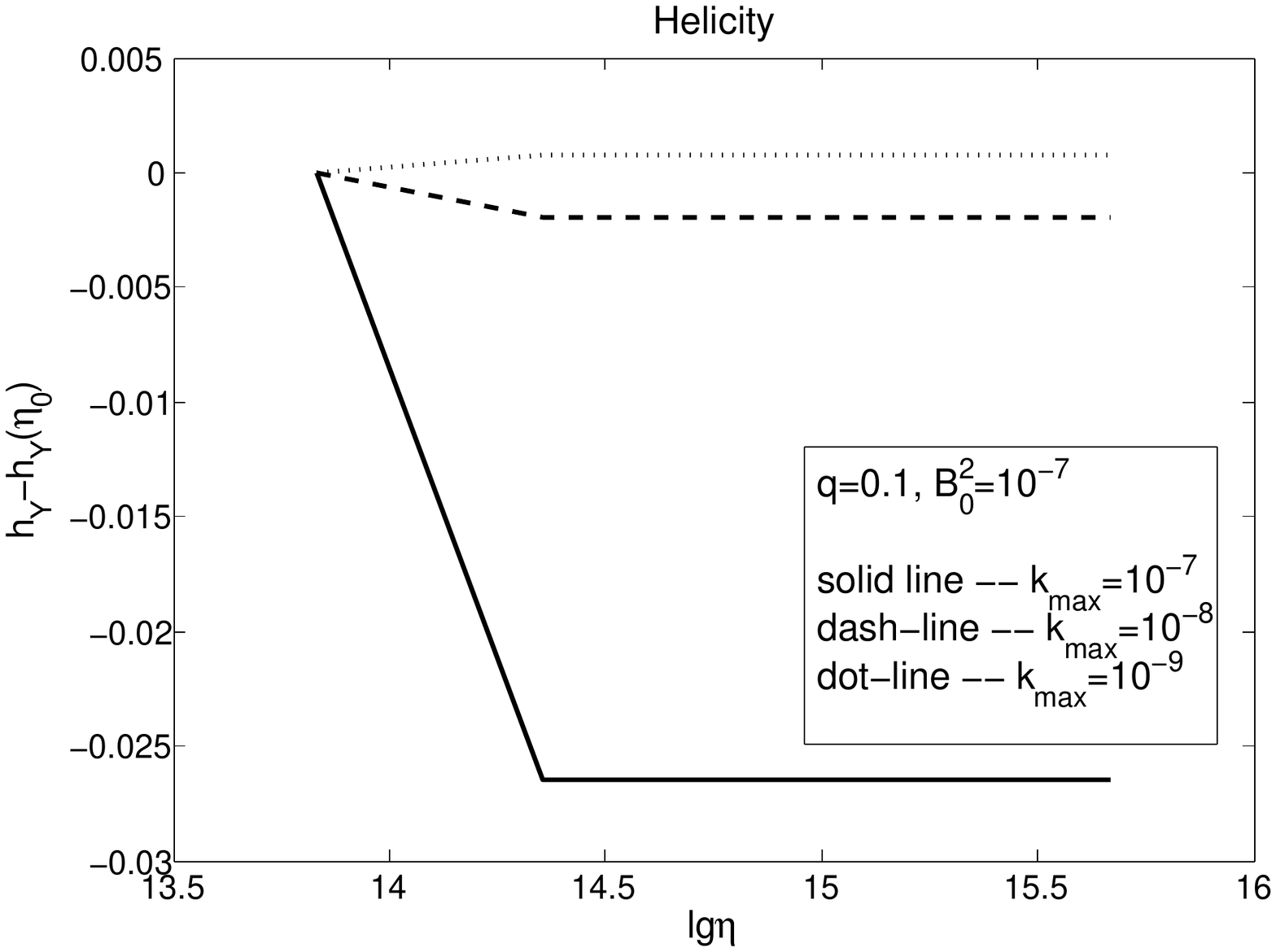}}
  
  \caption{Conservation of the HMF helicity density $\tilde{h}_Y(\eta)=\int d\tilde{k}\tilde{h}_Y(\tilde{k},\eta)$ for the Kolmogorov spectrum, $n_{B_Y}=-5/3$. Top panel corresponds to different $q=0.01,~0.1,~1$ and same conformal $\tilde{k}_{max}=k_{max}=10^{-6}$. 
Bottom panel shows a small distinction of the helicity density from the the initial one and corresponds to the Kolmogorov spectrum for different $k_{max}=10^{-9}\div 10^{-7}$ and same $q=0.1$. The seed (doubled) conformal HMF energy density equals to $2\rho_{B_Y}^{(0)}=B_0^2=10^{-7}$.}
  \label{fig:HDensity}
  \end{center}
\end{figure}
One can see in Fig. 3 that in the case of the Kolmogorov spectrum, $n_{B_Y}=-5/3$, the helicity density $\tilde{h}_Y(\eta)$ is almost conserved, $d\tilde{h}_Y(\eta)/d\eta\approx 0$, that is the consequence of the main contribution of large HMF scales \cite{footnote3} in the integral $\tilde{h}_Y(\eta)=\int_{k_{min}}^{k_{max}} d\tilde{k}\tilde{h}_Y(\tilde{k},\eta)$ near the horizon size at $\eta_0$, or at lower $\tilde{k}\sim \tilde{k}_{min}\simeq \tilde{l}_H^{-1}(\eta_0)\sim 10^{-14}$ . In such a case both the diffusion exponent and the argument in hyperbolic functions in Eq. (\ref{helicity_integral}) play a negligible role, so the helicity density $\tilde{h}_Y(\eta)$ depends rather on the parameter $q$ (when $q\cosh (...)\to q,~\sinh (...)\to 0$ ), or the spectrum is given mostly by its initial value Eq. (\ref{q-initial}), $\tilde{h}_Y(\eta_0)\approx 2q(\tilde{B}_0^Y)^2\tilde{k}_{min}^{-1}/5$. 

Nevertheless, a slight decrease of the helicity density due to $d\tilde{h}_Y/d\eta < 0$ (invisible in Fig. 3), that comes from the negative derivative of the diffusion exponent in Eq. (\ref{helicity_integral}) provides the growth of the right electron asymmetry $\xi_{eR}$ seen in Fig. 2 for different helicity levels $q$, or the helical HMF feeds the lepton asymmetry, cf. the first term in Eq. (\ref{right}). Note that such a derivative term in (\ref{right}), $d\tilde{h}(\tilde{k},\eta)/d\eta\sim - (2\tilde{k}^2/\sigma_c)\exp [- 2\tilde{k}^2(\eta - \eta_0)/\sigma_c]$, gives a convergence of the integral $\int d\tilde{k}\tilde{k}^{-2/3}(...)$ accounting for the factor $\sim \tilde{k}^{-8/3}$. Otherwise, the dependence on the upper limit value $\tilde{k}_{max}$ becomes more important  resulting in distinguishable curves in the bottom panel in Fig. 2 for running $\tilde{k}_{max}$ and fixed $q$.
Contrary to that, without such a differentiation $d\tilde{h}_Y/d\eta$ presented for the derivative $d\xi_{eR}/d\eta$ in  Eq. (\ref{right}), the bottom panel in Fig. 3 shows for $\tilde{h}_Y(\eta)=\int_{\tilde{k}_{min}}^{\tilde{k}_{max}} d\tilde{k}\tilde{h}_Y(\tilde{k},\eta)$ a slight dependence on varying upper limits $\tilde{k}_{max}$.

Note also that, in opposition to the case of the monochromatic spectrum illustrated in Fig. 1, the curves for HMF helicity density $\tilde{h}_Y(\eta)$ shown for different $q's$ in the top panel in Fig. 3 remain parallel, or do not tend the maximal helicity at $q=1$. Again this happens because the spectrum (\ref{helicity_integral}) is weighted by the first factor $2\tilde{\rho}_{B_Y}(\tilde{k},\eta_0)/\tilde{k}\sim \tilde{k}^{-8/3}$ for the Kolmogorov index $n_{B_Y}= -5/3$, so large HMF scales (small wave numbers) prevail in the integral $\tilde{h}_Y(\eta)=\int d\tilde{k}\tilde{h}_Y(\tilde{k},\eta)$. Really, even for the monochromatic spectrum, $\tilde{h}_Y(\tilde{k},\eta)\sim \delta (\tilde{k} - k_0)$, curves for a smaller $k_0$ have not a time to reach that maximum $\tilde{h}=2\tilde{\rho}_{B_Y}/k_0$ before EWPT, see the dotted curve ($k_0=10^{-8}$) in the top panel in Fig. 1.

\section{BAU evolution for the Kolmogorov HMF energy spectrum}
The BAU evolution given in our model by the Hooft's conservation law $B(t)/3 - L_e(t)=const$ \cite{footnote5}  where baryon number $B(t)$ "sits" in the HMF field,
\begin{eqnarray}\label{BAU1}
&&B(t)=3\int_{t_0}^t
\left[\frac{{\rm d}L_{eR}}{{\rm d}t^{'}} + \frac{{\rm d}L_{eL}}{{\rm d}t^{'}} +\frac{{\rm d}L_{\nu_{eL}}}{{\rm d}t^{'}}\right]dt^{'}\nonumber\\&&=\frac{3g^{'2}}{8\pi^2}\int_{t_0}^t({\bf E}_Y\cdot{\bf B}_Y)\frac{{\rm d}t^{'}}{s} - 3\int_{t_0}^t\Gamma_{sph}TL_{eL}dt^{'},
\end{eqnarray}
in conformal variables depends on the lepton asymmetries $\xi_{eR,eL}(\eta)$,
\begin{eqnarray}\label{BAU2}
B(\eta)=5.3&&\times 10^{-3}\int_{\eta_0}^{\eta}d\eta^{'}\Bigl[\frac{{\rm d}\xi_{eR}(\eta^{'})}{{\rm d}\eta^{'}} + \Gamma(\eta^{'})\Bigl(\xi_{eR}(\eta^{'}) \nonumber\\&& - \xi_{eL}(\eta^{'})\Bigr)\Bigr]- \frac{6\times 10^7}{\eta_{EW}}\int_{\eta_0}^{\eta}\xi_{eL}(\eta^{'})d\eta^{'}.
\end{eqnarray}
In Fig.~4 we show the BAU growth provided by the leptogenesis in HMF ($\sim d\xi_{eR}(\eta)/d\eta >0$). Note that for a large initial right electron asymmetry, $\xi_{eR}(\eta_0)=10^{-8}\div 10^{-6}$ seen in Fig.~2, the asymmetry derivative occurs negative, $\sim d\xi_{eR}(\eta)/d\eta <0$. This results in a dangerous antimatter production through Eq. (\ref{BAU2}), $B<0$. Therefore, our assumption for such a free parameter in our model seems to be excluded. Note also that this happens  when the Higgs decays prevail over the HMF feeding the leptogenesis through Abelian anomaly (via negative derivative $d\tilde{h}_Y/d\eta\sim - {\bf E}_Y\cdot{\bf B}_Y <0$  and corresponding positive HMF contribution in Eq. (\ref{right})). On the other hand, for a small initial asymmetry ($\xi_{eR}= 10^{-14}$ here) the HMF helicity contribution prevails over Higgs decays , so $d\xi_{eR}/d\eta$ is positive, and BAU reaches the observable value $B_{obs}\simeq 10^{-10}$ for some $\tilde{k}_{max}$ and a fixed $q$ in the top panel in Fig. 4, or for a fixed $\tilde{k}_{max}$ and some value $q$ of a helical HMF in the bottom panel. The bigger that helicity level $q\leq 1$ the sooner BAU grows.

\begin{figure}
  \centering
  %\subfigure[]
  {\label{2a}
  \includegraphics[scale=.34]{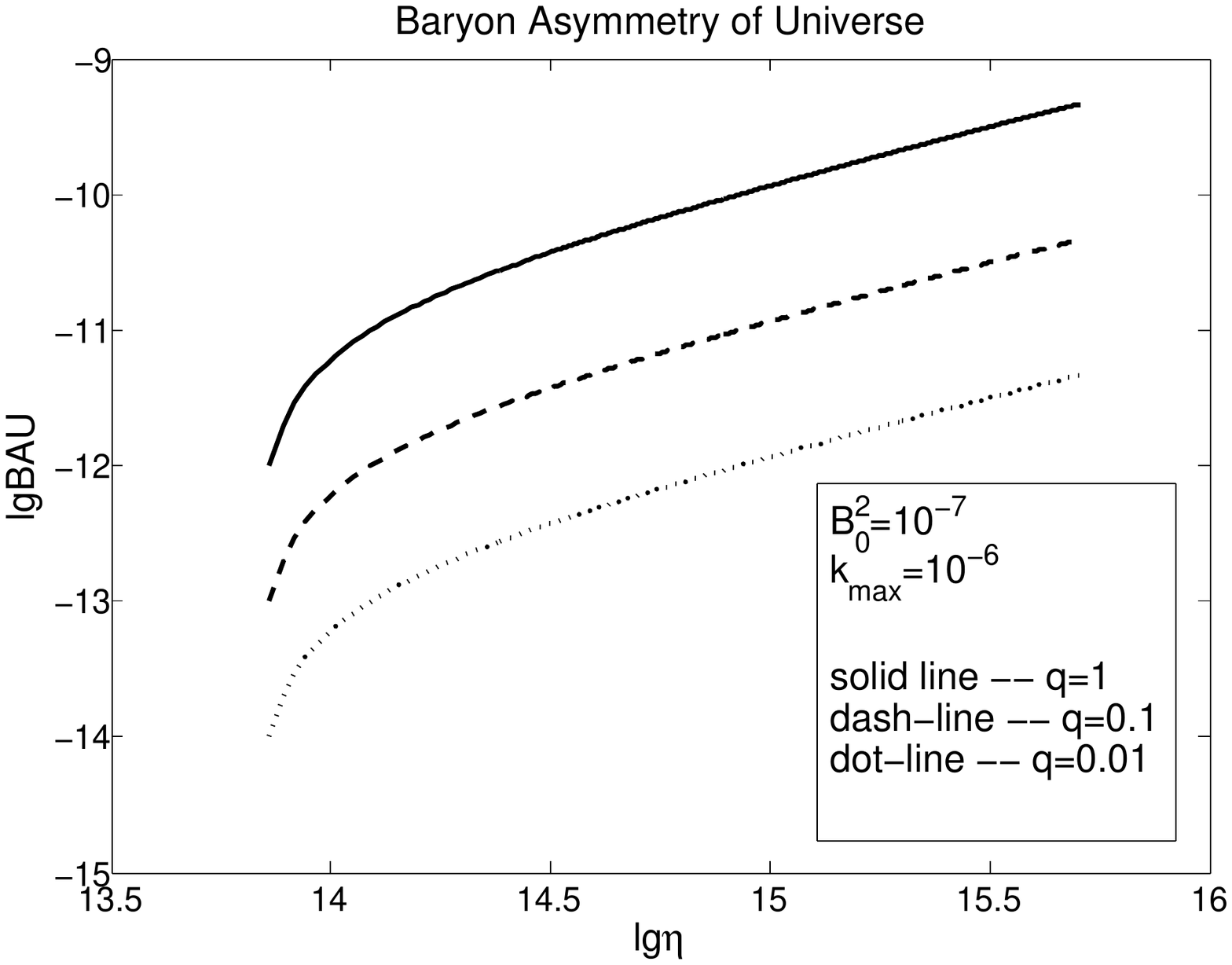}}
%  \hskip-.9cm
  %\subfigure[]
  {\label{2b}
  \includegraphics[scale=.34]{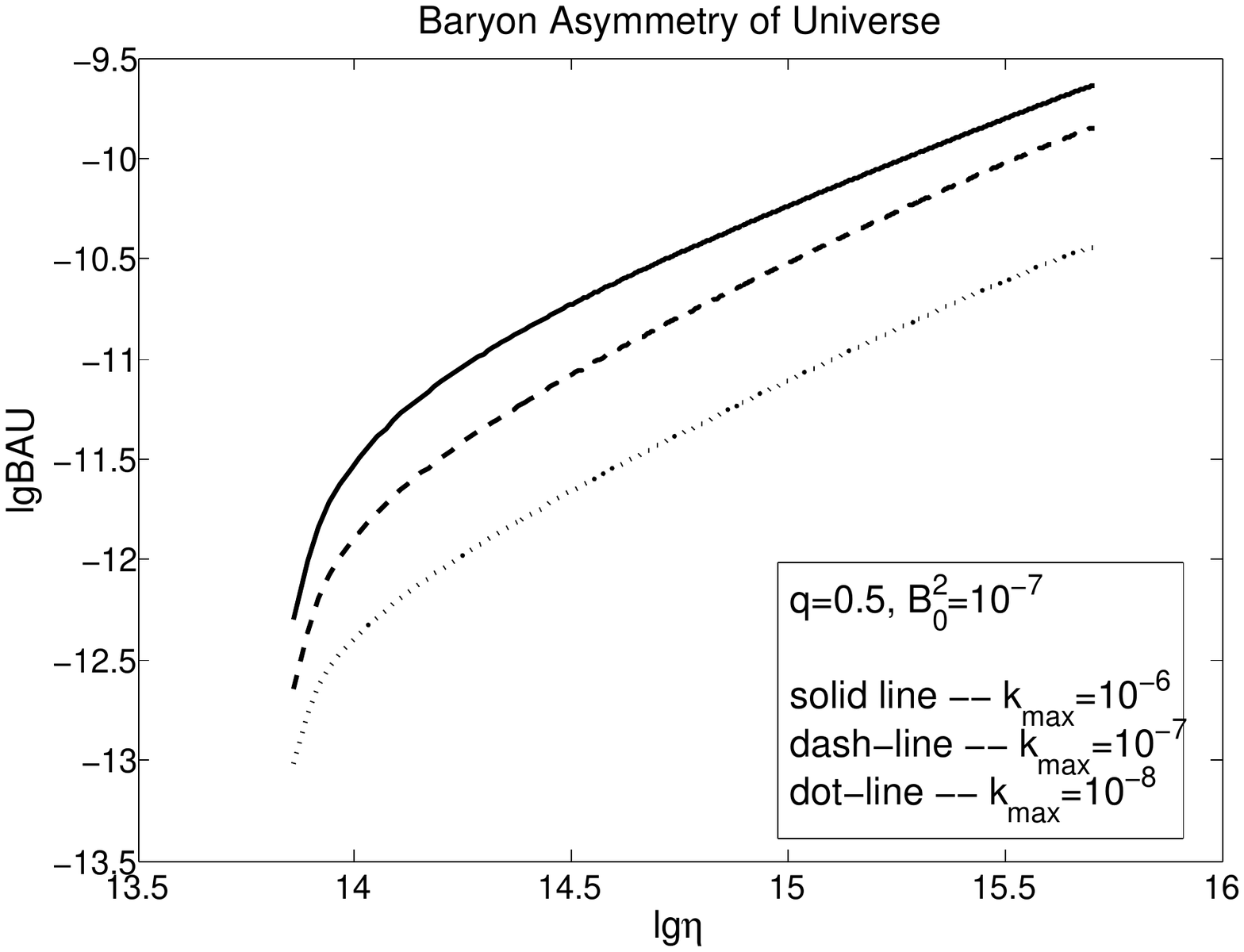}}
    \caption{The BAU evolution before EWPT in dependence on the conformal time $\eta$ for the Kolmogorov spectrum ($n_{B_Y}= -5/3$) given by the initial hypermagnetic energy density in Eq. (\ref{energy}).
 Top panel corresponds to the varying minimal scales of hypermagnetic fields, $\Lambda_{B_Y}=k_{max}^{-1}$, given by different 
$k_\mathrm{max}$ and  the fixed factor $q=0.5$ in Eq.(\ref{q-initial}).
 Bottom panel is built for the fixed $k_\mathrm{max}=10^{-6}$, and different levels of the initial hypermagnetic helicity in Eq.(\ref{q-initial}): $q=0.01$, $q=0.1$ and $q=1$. The doubled seed HMF energy density $B_0^2=10^{-7}$ is chosen as in Fig.3.
  \label{fig:BAU}}
\end{figure}

\section{Discussion}
We finish here by studying the simplest model for BAU generation based on the presence of an initial right electron asymmetry $\xi_{eR}(\eta_0)\neq 0$ \cite{Campbell:1992jd,Joyce:1997uy} in helical HMF. The model is based on the use of the two anomalies in SM: (i) the Chern-Simons one which appears due to parity violation and leads to the anomalous current (\ref{current}), and  (ii) the Abelian anomaly for lepton currents in HMF, see in Eq. (\ref{system}). The anomalous current (\ref{current}) which is added to the Ohmic one in Maxwell equation produces the $\alpha_Y$ -helicity parameter (\ref{alpha}) in Faraday equation (\ref{Faraday}), which, in turn, governs the HMF evolution. We considered the evolution of the spectra of binary products of hypercharge fields derived from Faraday equation: the HMF helicity density $h_Y(t)=V^{-1}\int d^3x ({\bf Y}\cdot{\bf B}_Y)$ and the HMF energy density  $\rho_{B_Y}(t)=V^{-1}\int d^3x ({\bf B}_Y^2)/2$. Assuming isotropic spectra in the Fourier representation, or in conformal (dimensionless) variables
$\tilde{h}_Y(\eta)=\int d\tilde{k}\tilde{h}(\tilde{k},\eta)$, $\tilde{\rho}_{B_Y}(\eta)=\int d\tilde{k}\tilde{\rho}_{B_Y}(\tilde{k},\eta)$, from corresponding kinetic equations (\ref{general}) we found the analytic relation of the HMF helicity spectrum $\tilde{h}(\tilde{k},\eta)$ with the lepton asymmetry $\Xi=\xi_{eR} + \xi_{eL}/2$ given by Eq.~(\ref{helicity_integral}). Then we completed our model considering the leptogenesis for such asymmetries $\xi_{eR}$, $\xi_{eL}$ given by the kinetic equations (\ref{right}), (\ref{left}) where we took into account the Higgs inverse decays and the weak sphaleron interaction with left lepton components in our model. Let us stress the crucial role of the HMF helicity for an efficiency of the leptogenesis and connected with it BAU generation.

In our simplified model we accounted for the weak sphaleron only which mediates a vacuum-vacuum transition in the $SU(2)_L$ sector and induces reactions among the weakly interacting particles (left electrons and left neutrinos) with the rate $\Gamma_{sph}\sim 25\alpha_WT$ .In this work, we checked the Kolmogorov spectrum to ensure that the left lepton 
asymmetry could not grow down to the EWPT time $\eta_{EW}$ washing out BAU through weak sphaleron processes, since asymmetry $\xi_{eL}$ remains much less than the right electron asymmetry, $\xi_{eL}\ll \xi_{eR}$, see Eq. (\ref{inequality}).
 The advantage of our model  is its simplicity comparing with more extensive approaches in some new works on the subject \cite{Long:2013tha,Long:2016uez,Fujita:2016igl}. In all works there is a common conclusion that for a more helical HMF the lepto/baryo -genesis proceeds more faster, and we demonstrated such issue here using MHD methods in apparent form for a realistic continuous (Kolmogorov) spectrum . 

Comparing plots in Fig. 1 (left panel) from our paper [6] and in Fig. 4 in the present work one can see both common BAU dependencies on time and some differences due to the use of different continuous spectra when relying on a more realistic arbitrary initial HMF helicity given by Eq. (9) in the present work instead of the maximum helicity spectrum $\tilde{k}\tilde{h}_Y(\eta, \tilde{k})=2\tilde{\rho}_{B_Y}(\eta,\tilde{k})$ used in our Ref. [6]. 
First, a big initial right electron asymmetry $\xi_{eR}(\eta_0)$ should be excluded in both cases:
both for the maximum helicity spectrum ( see Fig. 3 in paper [6] for $\xi_{eR}(\eta_0)= 10^{-4}$) and for an arbitrary HMF helicity here because of appearance of an antimatter production before EWPT, $B<0$. We did not show such a  danger negative BAU $B< 0$ (similar to the curve in Fig. 3 in [6]) referring below Eq. (18) to the plot in Fig. 2 for the negative derivative $d\xi_{eR}/d\eta < 0$ in the case of a large $\xi_{eR}(\eta_0)=10^{-6}$ that immediately leads to the negative sign for BAU evolution in Eq. (18). 
The growth of positive BAU with the increase of $k_{max}$ for a small initial $\xi_{eR}(\eta_0)=10^{-10}~{\rm in}~[6]~{\rm and}~\xi_{eR}(\eta_0)=10^{-14}$~here  is the common property for both HMF models (here and in Ref. [6]). Nevertheless, the observable $B_{obs}\sim 10^{-10}$ can be reached here for a larger $k_{max}$ : $k_{max}=10^{-7}\div 10^{-6}$ for $q=0.5$ in the top panel versus $k_{max}=10^{-9}\div 10^{-8}$ in the left panel in Fig. 1 ( Ref. [6]). This is because HMF helicity density $h_Y\sim YB_Y\sim kY^2$, which is proportional to the wave number and drives the lepto/baryogenesis, is weighted for the Kolmogorov spectrum by the factor $h_Y(k,\eta_0)\sim qB_Y/k\sim \tilde{k}^{-8/3}$ instead of the growing $h(k,\eta_0)\sim k^3$ used in our Ref. [6].

There was a recent work \cite{Sigl} where the modified MHD was considered in the both phases (symmetric and broken) around EWPT. In the symmetric phase the evolution of HMF energy ($\rho^Y_k$) and helicity ($h_k^Y$) spectra is governed by the right electron asymmetry $\mu_{eR}$ only in agreement with the approach \cite{Giovannini:1997eg} based on five global charges (correspondingly to five chemical potentials) conserved in SM. Our inclusion of the left asymmetry $\mu_{eL}$ was a necessary probe of a new equilibrium coming through Higgs decays below $T_0=10~{\rm TeV}$ when chirality flip processes enter equilibrium, $\Gamma_{RL}> H$, and we confirm validity of ideas by
authors \cite{Campbell:1992jd} that sphaleron processes in such a case are not danger , so $\mu_{eR}$ plays major role in given scenario. Note that authors \cite{Sigl} considered also the case of an arbitrary helicity for initial fields and obtained similar issues for magnetic helicity evolution above EWPT driven by the right electron asymmetry.
\section*{Acknowledgments}
We acknowledge Maxim Dvornikov for the useful discussions and Guenter Sigl and Natasha Leite for their remarks. D. S. is grateful to the RNF Grant No. 16-17-10097 for the financial support. A. S. is grateful to the financial support of the Ministry of Education and Science of the Russian Federation in the framework of Increase 
Competitiveness Program of MISiS.


\begin{thebibliography}{10}

\bibitem{Betal12}
A.~Brandenburg, D.~Sokoloff  and K.~Subramanian,
\textit{Current Status of Turbulent Dynamo Theory. From Large-Scale to Small-Scale Dynamos},
\textit{Sp. Sci. Rev.} \textbf{169}, 2012,  123-157.

\bibitem{Grasso:2000wj}
  D.~Grasso and H.~R.~Rubinstein,
  \textit{Magnetic fields in the early Universe},
  \textit{Phys. Rept.} \textbf{348} (2001) 163
  [astro-ph/0009061].
  

\bibitem{BK}
F. Krause, F., R. Beck, R., \textit{Symmetry and direction of seed 
magnetic fields in galaxies },
 \newblock  Astron. Astrophy., {\bf 335}, 789 (1998).

\bibitem{Neronov:2009gh}
  A.~Neronov and D.~V.~Semikoz,
  \textit{Sensitivity of gamma-ray telescopes for detection of magnetic fields in intergalactic medium},
  \textit{Phys. Rev.} \textbf{D 80} (2009) 123012
  [arXiv:0910.1920].
%%CITATION = 0910.1920;%%

\bibitem{Neronov:1900zz}
  A.~Neronov and I.~Vovk,
  \textit{Evidence for strong extragalactic magnetic fields from Fermi observations of TeV blazars},
  \textit{Science} \textbf{328} (2010) 73
  [arXiv:1006.3504].

\bibitem{Semikoz:2015wsa}
V.B. Semikoz and A. Smirnov , \textit{Leptogenesis in the Symmetric Phase of the Early
Universe: Baryon Asymmetry and Hypermagnetic Helicity Evolution},
\textit{J. Exp. Theor. Phys. } \textbf{120} (2015) 217-225 
[arXive:1503.06758].


\bibitem{Semikoz:2013xkc}
V.B. Semikoz, A.Yu. Smirnov and D.D. Sokoloff,
\textit{Hypermagnetic helicity evolution in early universe: leptogenesis and hypermagnetic diffusion},
\textit{JCAP} \textbf{10}	(2013) 014 [arXive:1309.4302].											

\bibitem{Akhmet'ev:2010ba}
  P.~M.~Akhmet'ev, V.~B.~ Semikoz and D.~D.~Sokoloff,
  \textit{Flow of hypermagnetic helicity in the embryo of a new phase in the electroweak transition},
  \textit{JETP Letters} \textbf{91} (2010) 215
  [arXiv:1002.4969].


\bibitem{Boyarsky:2011uy}
  A.~Boyarsky, J.~Fr\"{o}hlich and O.~Ruchayskiy,
  \textit{Self-consistent evolution of magnetic fields and chiral asymmetry in the early Universe},
  \textit{Phys. Rev. Lett.} \textbf {108} (2012) 031301 [arXive:1109.3350~[astro-ph]].

 
\bibitem{Dvornikov:2011ey}
  M.~Dvornikov and V.~B.~Semikoz,
  \textit{Leptogenesis via hypermagnetic fields and baryon asymmetry},
  \textit{JCAP} \textbf{02} (2012) 040
  [arXiv:1111.6876];
  \textit{Erratum:} \textit{JCAP} \textbf{08} (2012) E01.

\bibitem{Boyarsky:2012ex}
A.~Boyarsky, O.~ Ruchayskiy and M. Shaposhnikov,
\textit{Long-range magnetic fields in the ground state of the Standard Model plasma},
\textit{Phys. Rev. Lett.} \textbf{109} (2012) 111602
[arXiv:1204.3604 [hep-ph]] 

 
\bibitem{Semikoz:2009ye}
  V.~B.~Semikoz, D.~D.~Sokoloff and J.~W.~F.~Valle,
  \textit{Is the baryon asymmetry of the Universe related to galactic magnetic fields?},
  \textit{Phys. Rev.} \textbf{D 80} (2009) 083510
  [arXiv:0905.3365].


\bibitem{Semikoz:2012ka}
  V.~B.~Semikoz, D.~ Sokoloff and J.~W.~F.~Valle,
  \textit{Lepton asymmetries and primordial hypermagnetic helicity evolution},
  \textit{JCAP} \textbf{06} (2012) 008 
  [arXiv:1205.3607].


\bibitem{Dvornikov:2012rk}
M.~Dvornikov and V.~B.~Semikoz, 
\textit{Lepton asymmetry growth in the symmetric phase of an
electroweak plasma with hypermagnetic fields versus its washing out by sphalerons}
Phys. Rev \textbf{D87} (2013) 025023.

\bibitem{Boyarsky:2015faa}
A. Boyarsky, J. Frolich and O. Ruchayskiy, \textit{Magnetohydrodynamics of Chiral Relativistic Fluids},
Phys. Rev. \textbf{D92} (2015) 043004 [arXive: 1504.04854].

\bibitem{Shovkovy}
E. V. Gorbar, I. A. Shovkovy, S. Vilchinskii, I. Rudenok, A. Boyarsky, O. Ruchayskiy,
\textit{Anomalous Maxwell equations for inhomogeneous chiral plasma},
[arXive: 1603.03442]

\bibitem{Dvornikov:2013bca}
M. Dvornikov, and V.B. Semikoz, \textit{Instability of magnetic fields 
in electroweak plasma driven by neutrino asymmetries}, JCAP  \textbf{1405} (2014) 002
[arXive: 1311.5267].

\bibitem{Long:2013tha}
 A. Long and E. Sabancilar and T. Vachaspati,\textit{Leptogenesis and prinordial magnetic fields}, 
JCAP \textbf{1402} (2014) 036 [arXive: 1309.2315].



\bibitem{Fujita:2016igl}
T. Fujita, K. Kamada, \textit{Large-scale magnetic fields can explain baryon asymmetry of the Universe},
[arXive: 1602.02109].

\bibitem{Long:2016uez}
A. Long and E. Sabancilar, \textit{Chiral Charge Erasure via Thermal Fluctuations of Magetic Helicity},
[arXive: 1601.03777].

\bibitem{Semikoz:2011tm}
  V.~B.~Semikoz and J.~W.~F.~Valle,
  \textit{Chern-Simons anomaly as polarization effect},
  \textit{JCAP} \textbf{11} (2011) 048
  [arXiv:1104.3106].

\bibitem{Tashiro:2012mf}
Hiroyuki Tashiro, Tanmay Vachaspati and Alexander Vilenkin,
\textit{Chiral Effects and Cosmic Magnetic Fields},
\textit{Phys. Rev.} \textbf{D 86} (2012) 105033
[arXiv:1206.5549 [astro-ph.CO]].


\bibitem{Biskamp}
D. Biskamp, \textit{Magnetohydrodynamic Turbulence}, Cambridge University Press, Cambridge, 2003.

\bibitem{Rubakov}
D.~S.~Gorbunov and V.~A.~Rubakov,
  \textit{Introduction to the theory of the early Universe: Hot Big Bang theory},
  World Scientific Publishing Company, Singapore (2011), pg.~251.


\bibitem{Campbell:1992jd}
  B.~A.~Campbell, S.~Davidson, J.~Ellis and K.~A.~Olive,
  \textit{On the
  baryon, lepton-flavor and right-handed electron asymmetries of the
  universe}
  Phys. Lett. B \textbf{297} (1992) 118
  [hep-ph/9302221].

  
\bibitem{Tevzadze:2012kk}
A. Tevladze, L. Kisslinger, A. Brandenburg and T. Kahniashvili,
\textit{Magnetic fields from QCD phase transitions},
Astrophys.J. \textbf{759} (2012) 54 [arXive: 1207.0751 (2012)].

\bibitem{Joyce:1997uy}
M. Joyce and M. Shaposhnikov, \textit{Primordial magnetic fields, right-handed electrons, and
 the Abelian anomaly}, Phys. Rev. Lett. \textbf{79} (1997) 1193 [arXive:astro-ph/9703005].

\bibitem{Sigl}
P. Pavlovich, N. Leite and G. Sigl, \textit{Modified Magnetohydrodynamics around the electroweak phase transition},
[arXive: 1602.08419]

\bibitem{Giovannini:1997eg}
  M.~Giovannini and M.~E.~Shaposhnikov,
  \textit{Primordial hypermagnetic fields and triangle anomaly}
  Phys. Rev. \textbf{D 57} (1998) 2186
  [hep-ph/9710234].

\bibitem{Kahniashvili:2012uj}
T. Kahniashvili, A.G. Tevzadze, A. Brandenburg, and A. Neronov,
\textit{Evolution of Primordial Magnetic Fields from Phase
Transitions}, Phys. Rev. \textbf{D87} (2013) 083007 [arXive: 1212.0596].

\bibitem{footnote1}
Throughout the text we have neglected the bulk velocity evolution described by the Navier-Stokes equation since the length scale of the velocity variation $\lambda_v$ is much shorter than the correlation distance of the magnetic field, $\lambda_v\ll k^{-1}$, or infrared modes of the magnetic field  are practically unaffected by the velocity of plasma. In addition, the bulk velocity ${\bf v}$ does not contribute to the helicity evolution $dh_Y/dt\sim ({\bf E}_Y\cdot{\bf B}_Y)$ when the generalized Ohm law is substituted, ${\bf E}_Y= - {\bf v}\times {\bf B}_Y + \eta_Y\nabla\times {\bf B}_Y - \alpha_Y{\bf B}_Y$. A small scale $\lambda_v$ is also a reason why we omitted dynamo term $\nabla\times {\bf v}\times {\bf B}_Y$ in the Faraday equation.

\bibitem{footnote2}
In our causal scenario $\tilde{k}_{min}(\eta_0)=\eta_0^{-1}$ is the wave number at the initial time $\eta_0<\eta$. For the spectra which converge contrary to the Kolmogorov one we can put formally $\tilde{k}_{min}(\eta_0)=0$ at the lower limit.

\bibitem{footnote3}
The integrand in $\tilde{h}_Y(\eta)=\int d\tilde{k}\tilde{h}_Y(\tilde{k},\eta)$ is weighted by the first factor $\sim \tilde{k}^{-8/3}$ for the Kolmogorov spectrum, $n_{B_Y}=-5/3$, see Eq. (\ref{helicity_integral}).

\bibitem{footnote4}
We stress that we are dealing with helical fields and its spectrum can be rather more complicated than the classical Kolmogorov spectrum \cite{Kahniashvili:2012uj}.

\bibitem{footnote5}
See motivation for the choice of global charges in our paper \cite{Semikoz:2015wsa}.
\end{thebibliography}
\end{document}